\documentclass[journal]{IEEEtran}
\ifCLASSINFOpdf
\else
\fi

\usepackage{url}

\usepackage{amsmath}
\usepackage{graphicx}
\usepackage{multirow}
\usepackage{booktabs}

\usepackage{xcolor}
\newcommand{\ns}[1]{\textcolor{red}{[ns: #1]}}
\newcommand{\mv}[1]{{\textcolor{green}{[mv: #1]}}}
\newcommand{\wm}[1]{{\textcolor{cyan}{[wm: #1]}}}
\newcommand{\lgo}[1]{{\textcolor{blue}{[lgo: #1]}}}
\newcommand{\sz}[1]{{\textcolor{orange}{[sz: #1]}}}

\newcommand{\name}{DPU}
\newcommand{\trademark}{\textsuperscript{TM}}

\usepackage{soul}
\newcommand{\rev}[1]{#1}

\soulregister{\ns}{1}
\soulregister{\mv}{1}
\soulregister{\wm}{1}
\soulregister{\lgo}{1}
\soulregister{\sz}{1}

\begin{document}
%
\title{\name{}: DAG Processing Unit for Irregular Graphs with Precision-Scalable Posit Arithmetic in 28nm}
%
%
%

\author{Nimish~Shah, Laura~Isabel~Galindez~Olascoaga, Shirui~Zhao, Wannes~Meert,
         and~Marian~Verhelst,~\IEEEmembership{Senior~Member,~IEEE}%
\IEEEcompsocitemizethanks{\IEEEcompsocthanksitem N. Shah, S. Zhao and M. Verhelst are with the Department
  of Electrical Engineering - MICAS, KU Leuven, Belgium.\protect\\
  \IEEEcompsocthanksitem L. I. Galindez Olascoaga was with the Department of Electrical Engineering - MICAS, KU Leuven, Belgium. She is now with 
  the Department of Electrical Engineering and Computer Sciences, 
  University of California, Berkeley.\protect\\
  \IEEEcompsocthanksitem W. Meert is with the Department
  of Computer Science - DTAI, KU Leuven, Belgium.}
\thanks{© 2021 IEEE. Personal use of this material is permitted. Permission from IEEE must be obtained for all
other uses, in any current or future media, including reprinting/republishing this material for advertising
or promotional purposes, creating new collective works, for resale or redistribution to servers or lists, or
reuse of any copyrighted component of this work in other works. }
}

\maketitle


\vspace*{-8mm}

\begin{abstract}
Computation in several real-world applications like probabilistic machine learning, sparse linear algebra, and robotic navigation, can be modeled as irregular directed acyclic graphs (DAGs). The irregular data dependencies in DAGs pose challenges to parallel execution on general-purpose CPUs and GPUs, resulting in severe under-utilization of the hardware. This paper proposes \name{}, a specialized processor designed for the efficient execution of irregular DAGs. The \name{} is equipped with parallel compute units that execute different subgraphs of a DAG independently. The compute units can synchronize within a cycle using a hardware-supported synchronization primitive, and communicate via an efficient interconnect to a global banked scratchpad. Furthermore, a precision-scalable posit\trademark{} arithmetic unit is developed to enable application-dependent precision. The \name{} is taped-out in 28nm CMOS, achieving a speedup of 5.1$\times$ and 20.6$\times$ over state-of-the-art CPU and GPU implementations \rev{on DAGs of sparse linear algebra and probabilistic machine learning workloads}. This performance is achieved while operating at a power budget of 0.23W, \rev{as opposed to 55W and 98W of the CPU and GPU}, resulting in a peak efficiency of 538 GOPS/W with \name{}, \rev{which is 1350$\times$ and 9000$\times$ higher than the CPU and GPU, respectively}. Thus, with specialized architecture, \name{} enables low-power execution of irregular DAG workloads.

\end{abstract}

\begin{IEEEkeywords}
Graphs, irregular compute graphs, parallel processor, synchronization, precision, posit
\end{IEEEkeywords}

%
\IEEEpeerreviewmaketitle

\section{Introduction}
%
%
%
%
\IEEEPARstart{A}{directed} acyclic graph (DAG) is a directed graph with nodes and edges without cycles. DAGs are often used to model computation in programs, in which a node represents some compute operations, and the edges represent the dataflow and the dependencies among these operations.
For example, 
the computation of a fully-connected neural network can be represented as a DAG with nodes corresponding to neural operations, and edges corresponding to the connectivity among them, directed according to the dataflow.

\begin{figure}[!t]
\centering
\includegraphics[trim={0cm 0cm 0cm 0cm} , clip, width=\columnwidth]{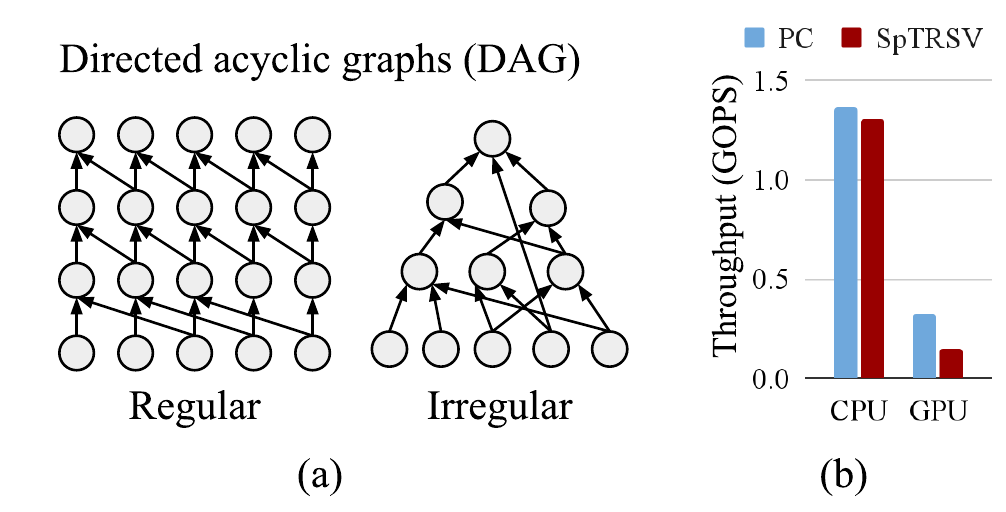}
\caption{(a) Regular (eg. dense linear algebra) vs irregular (eg. sparse linear algebra) DAGs. (b) Low GPU performance for irregular DAGs.}%
\label{fig:regular_vs_irregular}
\end{figure}

While the aforementioned neural network would result in a very regular DAG, several applications are characterized by highly \textit{irregular} DAGs, in which the edges point to seemingly-random nodes without repetitive patterns. For example, in a sparse neural network, around 60-70\% of the edges can be dropped \cite{parashar2017scnn}, leading to some irregularity in the DAG structure (fig. \ref{fig:regular_vs_irregular}(a)). Different applications exhibit different degrees of irregularity. In this work, we focus on highly-irregular DAGs resulting from more than 99\% sparsity in applications like sparse linear algebra, probabilistic machine learning, robotic localization, drone navigation, etc. \cite{dellaert2017factor,DBLP:conf/icml/StelznerPK19,8967568,suleiman2018navion}.

DAG irregularity poses several challenges for efficient hardware execution. For example, fig. \ref{fig:regular_vs_irregular}(b) shows CPU and GPU irregular DAG throughput corresponding to two applications: (1) probabilistic circuits (PC, also called sum-product networks) used in machine learning \cite{choi2020probabilistic,poon2011sum}, and (2) sparse matrix triangular solve (SpTRSV), a widely-used linear algebra operation \cite{davis2006direct}.
The GPU severely underperforms compared to the CPU despite having highly parallel hardware. Such a low performance 
is due to the irregularity. The seemingly-random edges are unsuitable for the commonly used single-instruction-multiple-data (SIMD) units and systolic arrays. Irregular edges also result in irregular memory accesses, leading to under-utilization of caches and memory bandwidth. These irregular accesses also prevent memory coalescing, which is crucial for high GPU performance. Moreover, parallelizing different parts of DAGs across multiple units (like CPU cores, GPU streaming multiprocessors, etc.) demands high communication and synchronization overhead.

To overcome the aforementioned challenges,
this paper proposes \textbf{\name{}}, a specialized DAG Processing Unit designed to efficiently execute highly-irregular computational graphs and provide an optimized solution for these emerging workloads. The \name{} is taped out in TSMC 28nm technology, and benchmarked on probabilistic machine learning and sparse linear algebra DAG workloads. Some of the key features are:
\begin{itemize}
    \item Parallel asynchronous compute units equipped with software-managed local scratchpads for data reuse, connected to a global banked scratchpad via a low-overhead \textit{asymmetric} crossbar for high memory bandwidth.
    \item Hardware support for fast synchronization across compute units, as frequently required for irregular DAGs.
    \item Execution based on decoupled data handling and compute streams to overlap memory and arithmetic instructions.
    \item Precision-scalable arithmetic units based on a customized posit\trademark{} representation to enable application-dependent precision selection.
\end{itemize}

The paper is organized as follows. The basics of irregular DAG execution and the related challenges are discussed in \S \ref{sec:background}. Section \S \ref{sec:DPU_architecture} describes the \name{} architecture and \S \ref{sec:compute_unit} explains the internal of a compute unit of the \name{}, followed with \S \ref{sec:posit_unit} that discusses the precision-scalable posit\trademark{} unit. Subsequently, \S \ref{sec:experiments} presents the experimental results and measurements of the taped-out processor. Finally, \S \ref{sec:related_works} and \S \ref{sec:conclusion} discusses the related work and concludes the paper.

\section{Background and Challenges} \label{sec:background}
\subsection{Background of DAG execution}
In this paper, a DAG represents a graph of nodes and edges encoding the computation to be performed for an application.
Depending on the application, the DAG nodes may represent one or a few arithmetic operations (e.g. addition, multiplications) on the node's inputs.
The output result of a node serves as an input to the
nodes connected to the outgoing edges of that node. Executing a computational DAG means evaluating the operations in all the nodes in a \textit{valid} order.
\subsubsection{\textbf{Execution order}}
The directed edges of a DAG represent data dependencies, which impose an ordering of execution of the nodes. For all the edges, the source node of an edge should be executed before the destination node. Put differently, a node can be executed only after all the \textit{predecessor} nodes connected to the incoming edges have finished execution. 

\subsubsection{\textbf{What can be executed in parallel?}}
At a given moment, all the nodes whose predecessor nodes have finished execution can be scheduled. These are called the \textit{active} nodes. The execution begins with the source
nodes of the DAG being \textit{active}, as they do not have any incoming edges. Subsequently, different nodes become active as the execution progresses. Note that the active nodes can be executed in parallel,
as there are no dependencies among active nodes by definition (otherwise they would not have been active together).

As such, DAG execution happens in multiple layers.
A set of nodes becomes active in each layer. The total number of layers is the same as the 
length of the 
longest path of nodes (critical path).
Parallelism
in a DAG can be quantified as the total number of nodes divided by 
the critical path length.
This metric quantifies a bound on the possible speedup over a sequential execution. Eg., a DAG with a parallelism of 100 cannot be executed faster than 100$\times$ even with infinite parallel units.
\subsubsection{\textbf{Synchronizations}}
If the active nodes in a layer are mapped across multiple \textit{asynchronous} compute units, then \textit{synchronization barriers} might be needed to maintain the required ordering before beginning a new layer of active nodes.
Examples of such asynchronous units are CPU cores and GPU streaming multiprocessors (also \name{}'s compute units, as explained later in the paper). Consider a DAG with 3 nodes as shown in fig. \ref{fig:synchronization_example}(a). Suppose nodes A and B of the first layer are scheduled on different units. To schedule node C from the next layer on unit 1, 
it must be ensured that node B has finished execution on unit 2 and its results are visible to unit 1.
But, by definition, asynchronous units do not provide such guarantees. Hence, an explicit synchronization barrier is needed before scheduling node C. 

\subsubsection{\textbf{Placement of nodes and synchronizations}} \label{sec:barrier_placement}
\begin{figure}[!t]
\centering
\includegraphics[trim={0cm 0.1cm 0cm 0cm} , clip, width=0.9\columnwidth]{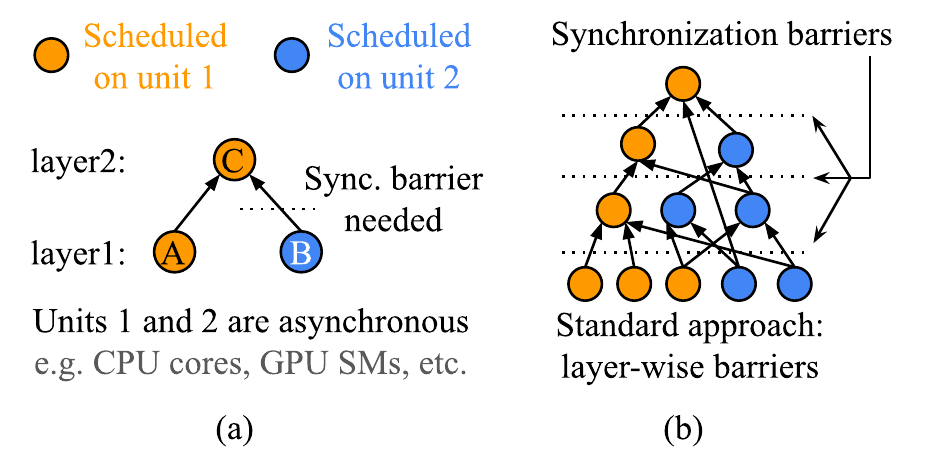}
\caption{\textbf{Synchronizations.} (a) Need for a synchronization in a simple DAG. (b) Commonly-used \textit{layer-wise} synchronization placement}%
\label{fig:synchronization_example}
\end{figure}

For applications such as probabilistic machine learning and sparse linear algebra, the DAG structure is fully known at compile time, allowing strategic mapping of nodes to the compute units and the placement of the synchronizations by a compiler. The aim is to reduce the total number of synchronizations and the associated overhead. Fig. \ref{fig:synchronization_example}(b) shows the standard approach of \textit{layer-wise} synchronizations (also called level scheduling), first introduced in \cite{10.1142/S0129053389000056}, where a synchronization is used after every layer of active nodes. The active nodes in a layer are mapped across available parallel compute units that then have to be synchronized before executing the next layer.

\begin{figure}[!t]
\centering
\includegraphics[trim={0cm 0cm 0cm 0cm} , clip, width=0.8\columnwidth]{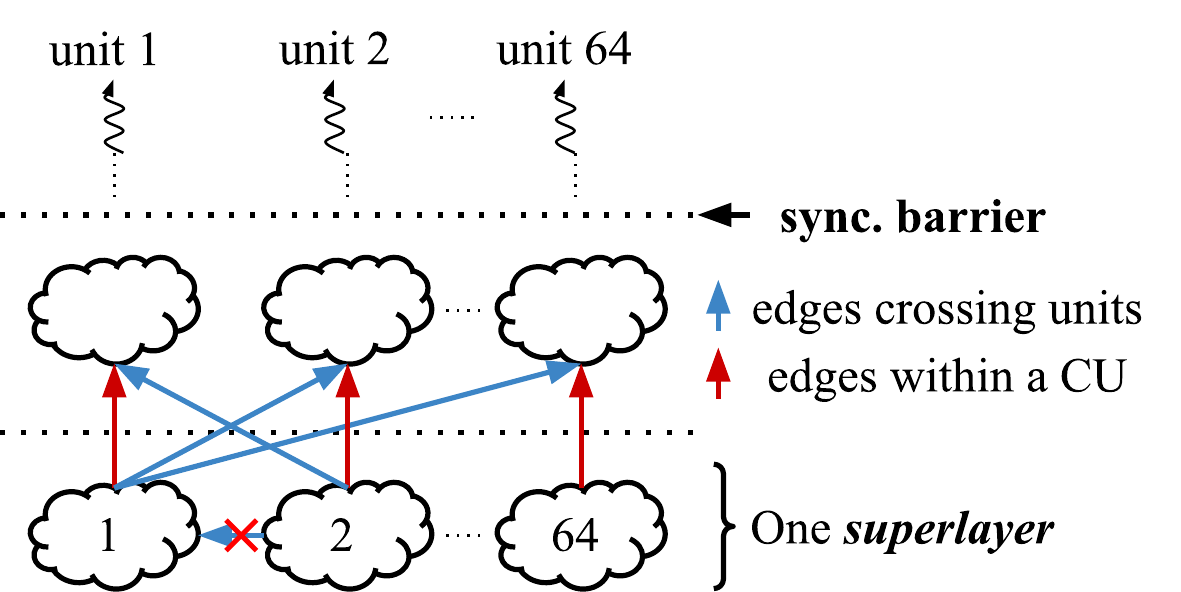}
\caption{Our approach of \textbf{\textit{superlayer}}-based synchronization placement}%
\label{fig:superlayers}
\end{figure}

\name{} uses
an advanced graph-partitioning technique to reduce the number of synchronization barriers, as explained in \cite{shah2021graphopt}. The DAGs are partitioned into \textit{superlayers} (fig. \ref{fig:superlayers}), each containing 64 subgraphs (possibly spanning multiple DAG layers) for 64 compute units in \name{}. A synchronization barrier is used after every superlayer. The subgraphs are made as large as possible to reduce the number of synchronizations, but also optimized to have similar sizes to balance the workload. This is achieved by translating the DAG partitioning task into a constrained-optimization problem and using the Google OR-Tools solver \cite{ortools}. As such, given a DAG, the mapping of nodes to the compute units and the placement of synchronizations are fixed at compile time.
\subsection{Challenges due to irregularity}
The irregularity in DAG structure poses several challenges for efficient execution on general-purpose processors like CPUs and GPUs. These are summarized in table \ref{tab:challenges} along with the related \name{}'s innovations to address them.
\subsubsection{\textbf{SIMD unfriendly}} 
The active nodes to be executed in parallel may perform the same arithmetic operation on the inputs, which would make them suitable for a SIMD execution. However, the inputs of these nodes typically reside in random memory locations owing to the irregularity of the DAG structure, making them unlikely to be co-located in a CPU or GPU cache-line. In fact, some of them may not be cached at all and need to be fetched from external memory. Experiments in \cite{8573480} find that around 50\% of the load requests in graph workloads result in cache misses. This leads to high variability in the load latency of these inputs, causing all the SIMD lanes to stall for the slowest input. 
Thus, despite the availability of parallel operations to execute, random memory loads make irregular DAGs SIMD-unfriendly. 
\renewcommand{\arraystretch}{1.2}
\begin{table}[!t]
\centering
\caption{Challenges and opportunities in irregular DAG execution and related \name{} innovations
}
\begin{tabular}{ll}
\toprule
\multicolumn{1}{c}{\begin{tabular}[c]{@{}c@{}}\textbf{Challenges/opportunities}\\ \textbf{for irregular DAGs}\end{tabular}} & \multicolumn{1}{c}{\textbf{\name{} innovations}} \\ \midrule
SIMD unfriendly & \begin{tabular}[c]{@{}l@{}}Asynchronous compute units \\ with independent instructions\end{tabular} \\  \hline
Frequent synchronizations & \begin{tabular}[c]{@{}l@{}}Hardware-supported synchronization \\ with special instructions, and \\\textit{superlayer}-based parallel execution \end{tabular} \\ \hline
Inefficient use of caches & Software-managed scratchpads \\ \hline
Data prefetching & \begin{tabular}[c]{@{}l@{}}Decoupled-instruction streams for \\ efficient hardware prefetching\end{tabular} \\ \hline
\begin{tabular}[c]{@{}l@{}}
Diverse applications with \\ varying precision requirement \end{tabular} & 
\begin{tabular}[c]{@{}l@{}}
Precision-scalable custom posit\trademark{} \\ arithmetic \end{tabular} \\
\bottomrule
\end{tabular}%
\label{tab:challenges}
\end{table}

Consequently, the x86 CPU SIMD vector instructions are not useful for irregular DAGs, and parallelizing threads across multiple CPU cores are preferred from a performance point of view. GPUs, on the other hand, can tolerate irregular memory latency by switching thread warps when a thread stalls (given there are enough thread warps available to schedule). However, GPUs suffer from other bottlenecks as discussed subsequently. The \name{} uses asynchronous compute units that execute independent instruction streams instead of a SIMD unit.

\subsubsection{\textbf{Frequent synchronizations}} \label{sec:frequent_sync}
The total amount of computation done per synchronization barrier is a key indicator of parallel performance --- the higher the better, to amortize the synchronization cost. This amount depends on the DAG structure and the barrier-placement techniques (refer \S \ref{sec:barrier_placement}). In our benchmarks, the \textit{layer-wise} approach results in 210 compute operations per barrier, which increases to 633 with the superlayer-based approach for 64 parallel units. Yet, if the barrier takes comparable cycles as the computations, it can severely degrade the parallel performance, possibly making it even worse than a sequential execution.

The synchronization barriers in multi-core CPUs are implemented with atomic operations on a shared memory location. These atomic operations typically incur long latency from 
the CPU core to the outer-most shared caches or external memory. Furthermore, these operations also create a burst of cache-coherency traffic as every core modifies the same location, increasing the barrier cost further. Due to these reasons, the synchronization barriers in CPUs typically take 3000 cycles (measured with the EPCC microbenchmark \cite{bull2012microbenchmark}). Similarly, in the GPUs, the global synchronization for all the CUDA cores consumes around 2000 cycles \cite{zhang2020study}. To avoid this bottleneck, \name{} is equipped with a hardware-supported barrier instruction that synchronizes all the units in a single cycle.

\subsubsection{\textbf{Inefficient use of caches}} DAG execution results in randomly addressed single-word memory accesses (typically of 4B) with lower spatial locality, which implies that the 32/64 words cacheline granularity of CPU and GPU is too high, and most of the fetched words are unlikely to be used. Such random accesses also prevent the memory-request coalescing,
which is critical for good GPU performance. Smaller cachelines are preferable for these workloads, but lead to higher area/energy overhead of tag storage and lookup. 
Furthermore, depending on the access patterns, some words are reused much more frequently than the others. Thus, selectively choosing which words to store locally results in an efficient use of the scarce on-chip local storage. 
Due to these reasons, \name{} uses software-managed local scratchpads with single-word accesses, instead of hardware-managed caches. 

\subsubsection{\textbf{Data prefetching}} 
An out-of-order CPU core tries to find the data-prefetching opportunities at runtime, and can issue multiple out-standing load requests \cite{hammarlund2014haswell, doweck2017inside}. However, \cite{beamer2015locality} reports that an Intel CPU could only keep 2-3 load requests in flight while executing graph workloads, even when the architecture supported up to 10 load requests. The main reason for this underperformance is the limited instruction window in which the core looks for independent load requests, and widening this window is area and power-hungry. To prefetch data efficiently, \name{} exploits the fact that the DAG structure is known at compile time. The compiler decouples the \name{}'s instructions into memory and processing \textit{streams}, which are executed independently on different subunits. This enables low-overhead prefetching and memory-compute overlap, without using costly out-of-order hardware.

\section{\name{} architecture} \label{sec:DPU_architecture}
\subsection{Compute units (CU)}
\begin{figure}[!t]
\centering
\includegraphics[trim={0cm 0cm 0cm 0cm} , clip, width=0.9\columnwidth]{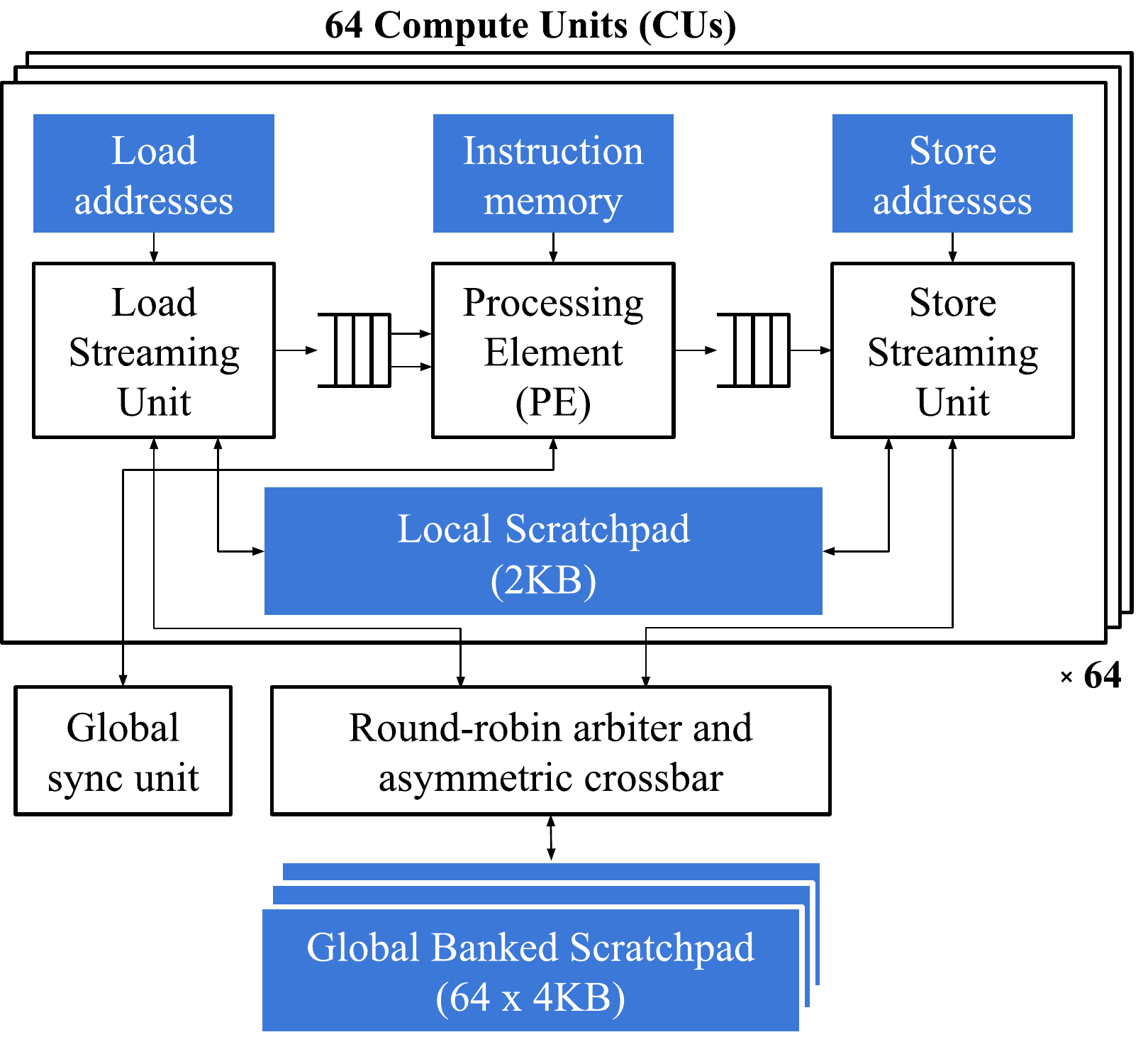}
\caption{The \textbf{\name{} architecture} with 64 parallel CUs}%
\label{fig:full_architecture}
\end{figure}

Fig. \ref{fig:full_architecture} shows the DPU architecture with 64 parallel \textbf{compute units (CU)} that execute 64 subgraphs in the superlayers shown in fig. \ref{fig:superlayers}.
Each CU is equipped with its own instruction memory and executes these instructions \textit{asynchronously}; a stalled CU does not stall the others. The CUs communicate via a global scratchpad connected by an asymmetric crossbar (\S \ref{sec:global_scratchpad}).
The single-cycle synchronization of CUs is made possible with a \textit{global barrier} instruction and a global sync unit (\S \ref{sec:global_sync_unit}).
The detailed architecture of CU is explained later in \S \ref{sec:compute_unit}. A specialized compiler \cite{shah2021graphopt} is designed that takes an arbitrary DAG and generates the superlayers, schedules operations, performs memory allocation, etc. for \name{} execution.

\subsection{Global scratchpad and asymmetric crossbar} \label{sec:global_scratchpad}
As observed in \cite{8573480}, irregular graphs typically lead to relatively high global traffic when executed on parallel units. In our benchmarks, we also observe a similar trend. The blue global edges in superlayers in fig. \ref{fig:superlayers} accounts for 33\% of the total edges in our benchmarks. In \name{}, this global inter-CU communication happens via a high-bandwidth global scratchpad. The 256KB scratchpad is constructed with 64 banks of 4KB each, providing an overall bandwidth of 2Kb/cycle. 

\begin{figure}[!t]
\centering
\includegraphics[trim={0cm 0cm 0cm 0cm} , clip, width=0.75\columnwidth]{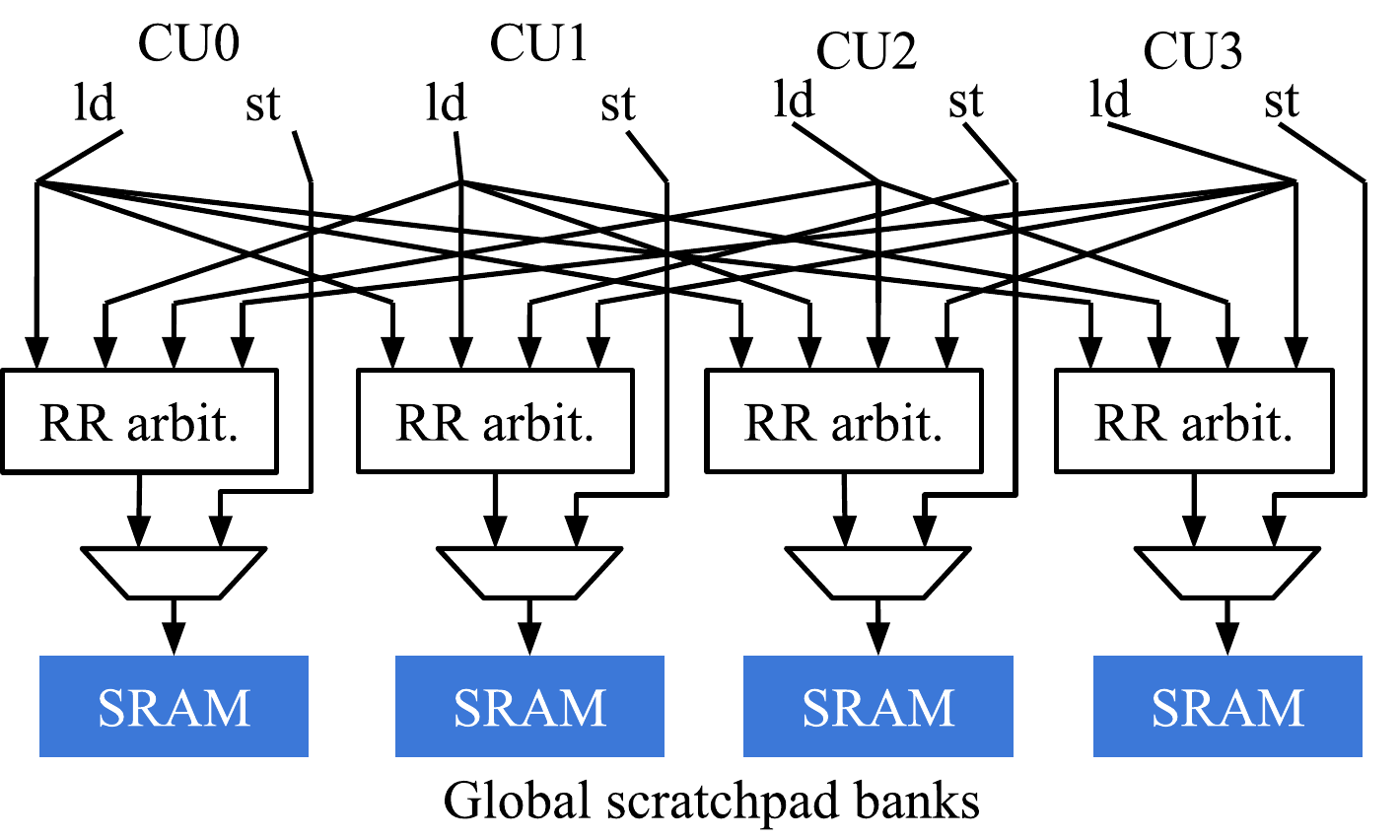}
\caption{\textbf{Asymmetric crossbar} to reduce area and power overhead}%
\label{fig:asymmetric_crossbar}
\end{figure}%

Furthermore, for low hardware overhead, the CUs connect to the global scratchpad via an \textit{asymmetric} crossbar (fig. \ref{fig:asymmetric_crossbar}), such that a CU can load data from any bank but can store to only one specific bank each. This asymmetry of global loads but restricted stores (instead of the other way around) is a deliberate design choice considering the fact that an output of a node is stored only once but usually loaded multiple times from different CUs. 

\rev{The asymmetric design does reduce the flexibility of mapping intermediate node outputs to the global scratchpad banks. In fact, the bank mapping is fully determined based on the mapping of operations to the CUs (which happens during the superlayer generation), because the output of an operation on a CU can only be stored to that CU's respective bank. On the other hand, with a full crossbar, the compiler could possibly map an output to any bank to reduce bank conflicts. However, predicting and averting these bank conflicts during compilation is anyway not possible given that the CUs are asynchronous and can possibly stall for unpredictable cycles (for example, waiting for the next instruction). As a result, in practice, it is very difficult for a compiler to exploit the additional flexibility coming from a full crossbar, and hence the inflexibility of an asymmetric crossbar does not drastically impact the throughput.
}

For each bank, the load requests are selected based on a round-robin arbitration scheme,
while the store request has the highest priority. The store request cannot participate in the round-robin arbitration with 64 load requests because that would reduce the store bandwidth by 64$\times$ compared to the loads. 
Overall, this asymmetric crossbar consumes 45\% lower area and energy than the symmetric counterpart that allows store requests to reach every bank.

\subsection{Global sync unit} \label{sec:global_sync_unit}
To mitigate the overhead of frequent synchronizations, the CUs are equipped with a special instruction for global barrier, complemented with a central dedicated global sync hardware unit. When a CU reaches a global barrier instruction, it indicates this to the global sync unit and stalls until the other CUs hit their global barrier instruction. The global sync unit uses a tree of AND gates to determine if all the CUs have reached the barrier, and communicates this to all the CUs within that cycle, enabling a single-cycle synchronization. Even though the paths to/from the global sync units create a long combinational loop due to the unit's centralized role, they are not the critical paths in our design. 



\section{Compute unit (CU) architecture}\label{sec:compute_unit}
\subsection{Local scratchpad}\label{sec:local_scratchpad}
Because the subgraphs in the superlayers are made as large as possible, they have a significant proportion of intra-CU edges shown in red in fig. \ref{fig:superlayers} (67\% of the total number of edges in our benchmarks). To exploit this locality, the CUs are equipped with an address-mapped local scratchpad. The compiler maps the output data of a node to the local scratchpad if all outgoing edges of the node are intra-CU (colored red in fig. \ref{fig:superlayers}), otherwise it is mapped to the global scratchpad. A scratchpad is used instead of a cache due to the following reasons:   
\begin{itemize}
    \item A software-managed scratchpad can selectively store only the data that has local reuse.
    \item The typical cacheline granularity of 32 or 64 words is too large for graphs due to frequent irregular memory fetches, and results in wasted interconnect traffic \cite{8573480, beamer2015locality}.
    \item An address-mapped scratchpad avoids tag storage and lookup, reducing the area and energy footprint.
\end{itemize}

\begin{figure}[!t]
\centering
\includegraphics[trim={0cm 0cm 0cm 0cm} , clip, width=\columnwidth]{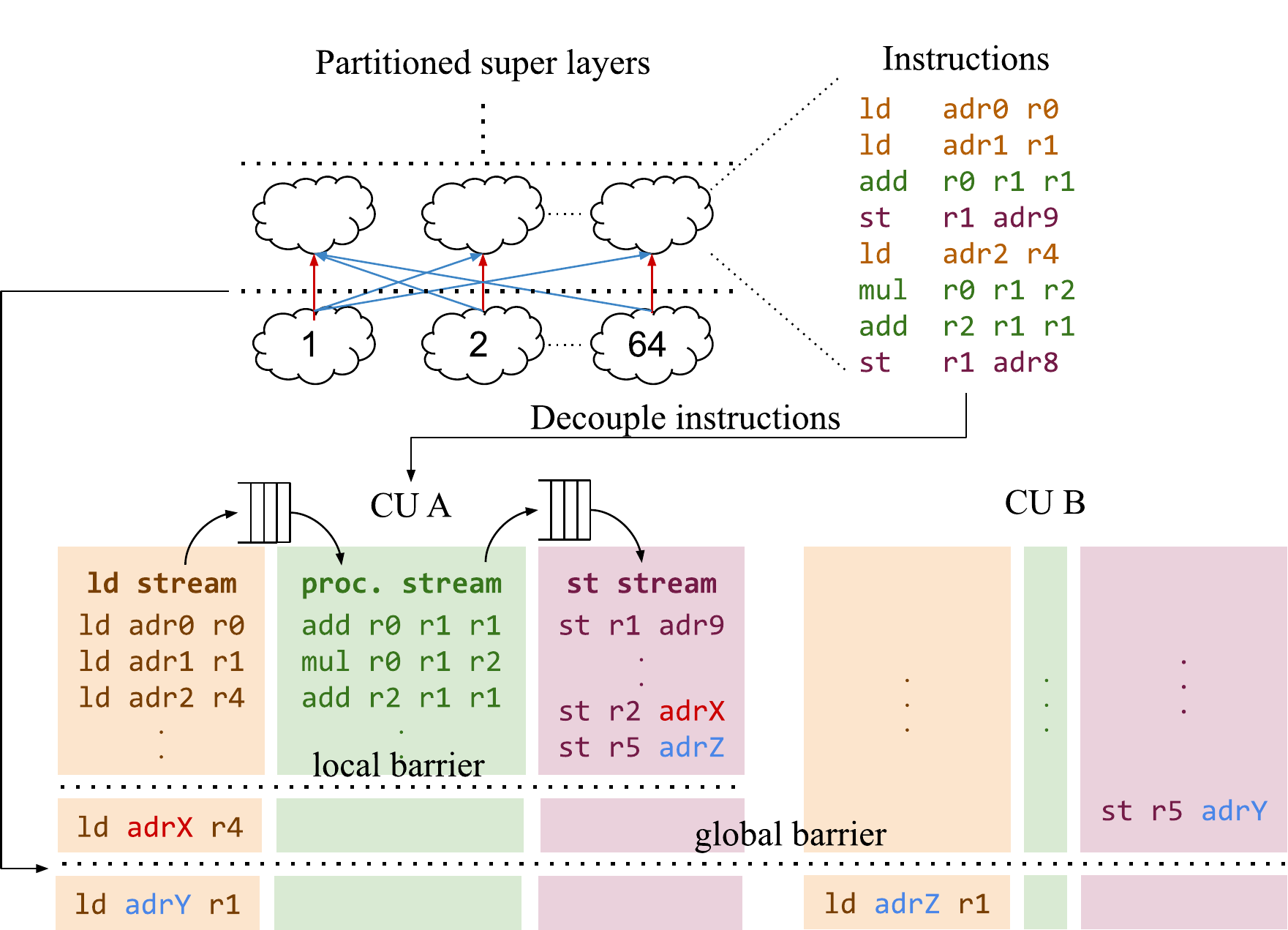}
\caption{\textbf{Decoupled streams} to overlap memory and processing instructions}%
\label{fig:decoupled_streams}
\end{figure}%

\subsection{Data prefetching using decoupled instruction streams} \label{sec:decoupled_streams}
For a given DAG, the memory load and store instructions to/from the scratchpads can be predicted at compile time. This is
leveraged to perform aggressive data prefetching by overlapping memory requests with arithmetic operations without the need for expensive out-of-order hardware. As shown in fig. \ref{fig:decoupled_streams}, the instructions for CUs are decoupled into three \textit{streams}: load, processing and store streams, which are executed independently on different components of the CU.

\textbf{Load streaming unit}: 
The load addresses (for both global and local scratchpads) along with the corresponding destination registers are programmed in the load address memory. The load streaming unit prefetches the data by issuing these load requests to the local or global scratchpads. The loaded data may not yet be allowed to be written to the register file in the PE, because the PE might still be using the destination register for some other computation. Instead, the data is pushed on a FIFO going to the PE, from where the PE will eventually consume it at the right moment. The load streaming unit keeps streaming the load requests as long as there is space in the FIFO. The prefetching cannot cross barriers, hence the length of the stream is controlled by a special \textit{load stream length} register, which is programmed by the PE after every barrier. The FIFO becomes empty at the barriers,
hence frequent barriers reduce prefetching efficiency. 

\begin{figure}[!t]
\centering
\includegraphics[trim={0cm 0cm 0cm 0cm} , clip, width=\columnwidth]{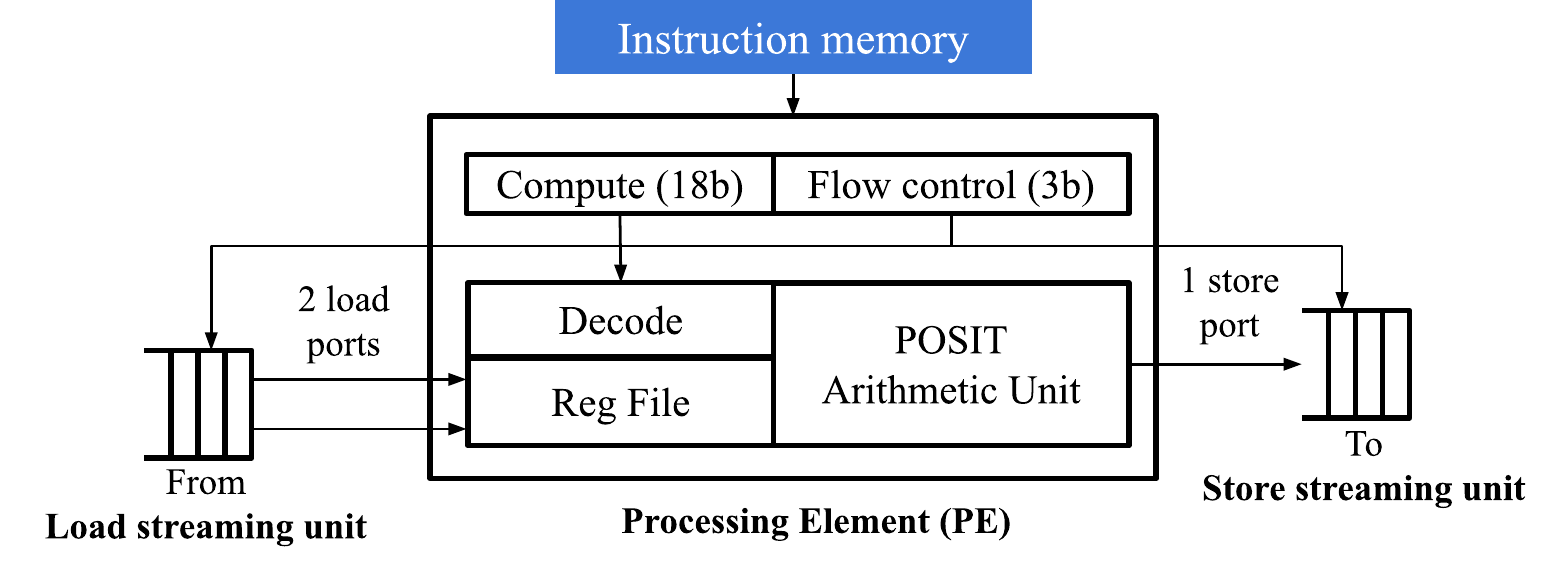}
\caption{\textbf{Internal PE design} and FIFO flow-control for precise ld/st timing}%
\label{fig:PE_architecture}
\end{figure}

\textbf{PE}: The processing streams contain the actual arithmetic instructions to be executed on the PE (fig. \ref{fig:PE_architecture}). A custom instruction set is designed (table \ref{tab:isa}), which dictates the computation of the arithmetic unit in the PE. \rev{The PE does not contain pipeline stages, and all the instructions have a latency of 1 cycle.} The instructions have an 18b compute field, with a 3b opcode for ALU and special-function instructions (table \ref{tab:isa}) and 3$\times$5b operand register file addresses. A 32-entry 32b register file is used with 3 write ports (2 for load ports and 1 for the arithmetic unit) and 2 read ports (for the arithmetic unit). The compiler makes sure that the 3 write ports write to different registers in the same cycle, to avoid conflicts. This is done by precisely controlling the timing of the loads, using flow control bits (see further). The output of the arithmetic unit connects to one of the register write ports and the FIFO to the store streaming unit.

The 3 remaining instruction bits control the PE's IO dataflow. The PE cannot directly communicate with scratchpads; it only gets/puts data from/to the FIFOs of load/store units. The FIFO flow control bits are encoded in the processing instructions, to let the PE precisely control the timing of migrating data from the load FIFO to the register file, resp. from the ALU output to the store FIFO. These flow control bits indicate whether a FIFO load/store is needed in parallel with the compute operation of the current instruction. The PE stalls if a load flow control bit is set, but there is no data in the load FIFO, or if the store bit is set but the store FIFO is full. With such a flow control, PE avoids data hazards like the write-after-read (WAR) hazard in which the data from load FIFOs overwrite an active register that is not yet fully consumed. 

\renewcommand{\arraystretch}{1}
\begin{table}[]
\centering
\caption{Instructions of PE}
\begin{tabular}{ll}
\toprule
\texttt{add, mul} & Add or multiply two numbers \\
\texttt{max, min} & \begin{tabular}[c]{@{}l@{}}Max or min of two numbers\\\end{tabular} \\
\begin{tabular}[c]{@{}l@{}} \texttt{global} or\\ \texttt{local barrier} \end{tabular} 
& Barriers for synchronization \\
\texttt{set\_ld\_stream\_len} & \begin{tabular}[c]{@{}l@{}}Sets the load stream length until next barrier\\ \end{tabular} \\
\texttt{set\_precision} & \begin{tabular}[c]{@{}l@{}}Sets the precision of the arithmetic unit\\ \end{tabular} \\ \bottomrule
\end{tabular}%
\label{tab:isa}
\end{table}

\begin{figure}[!b]
\centering
\includegraphics[trim={0cm 0cm 0cm 0cm} , clip, width=\columnwidth]{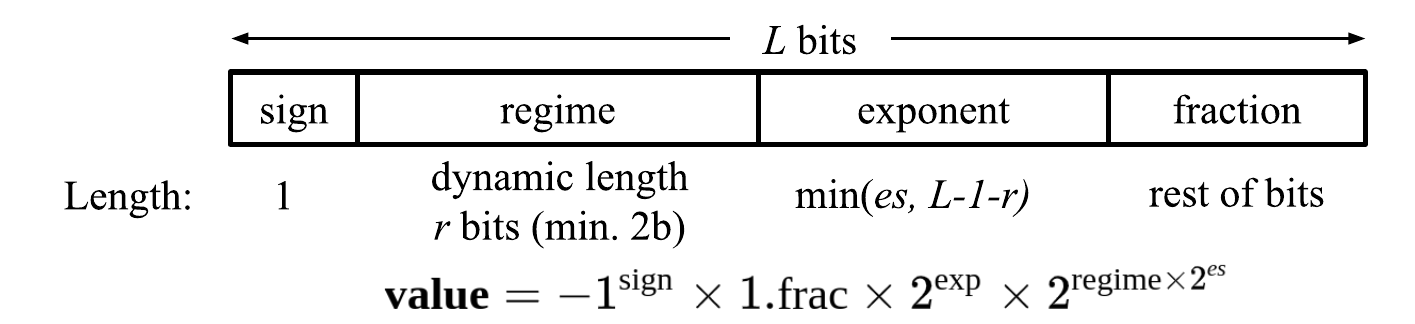}
\caption{The posit\trademark{} representation}%
\label{fig:post_representation}
\end{figure}%

\begin{figure*}[!t]
\centering
\includegraphics[trim={0cm 0cm 0cm 0cm} , clip, width=\textwidth]{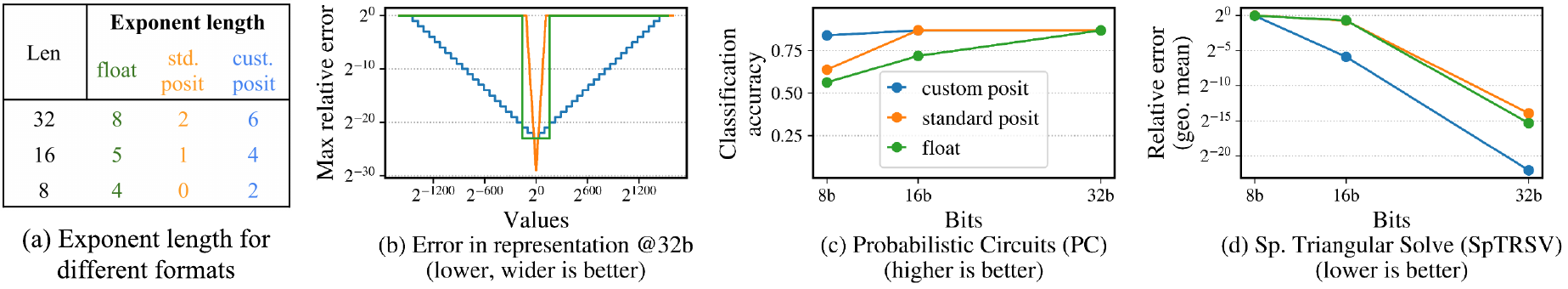}
\caption{\textbf{Accuracy impact of different representations}. Custom posit performs better due to lower error for wider range of values.}%
\label{fig:posit_accuracy_impact}
\end{figure*}%

\textbf{Store streaming unit}: The store streaming unit waits for data to show up on the store FIFO from the PE, and stores them to local/global scratchpads according to the addresses in the store stream. It also informs the PE whether all the data is stored to the memory and there are no outstanding store requests left, which helps the PE to decide if the compute unit is ready to sync with other units at the barriers.

\textbf{Local barrier.} The load-to-PE and the PE-to-store communication happens via FIFOs, and the corresponding data-dependencies are resolved by the FIFOs' flow control. There is also a dependency of load stream on the store stream. A load following a store to the same scratchpad address (adrX in fig. \ref{fig:decoupled_streams}) should not be executed before the store finishes. Since the streaming units operate independently, this ordering is not guaranteed. To address this, an intra-CU \textit{local barrier} is used. The \textit{load stream length} is programmed such that the load streaming unit waits at the local barrier for the store unit to catch-up before proceeding. 

\rev{\textbf{Stream generation}: Given a subgraph to be executed on a CU, the \mbox{\name{}} compiler first generates the corresponding instructions, by (1) scheduling the operations of subgraph nodes in a topological order using a depth-first traversal, (2) performing register allocation using the linear-scan method \mbox{\cite{poletto1999linear}}, and (3) inserting the load-store instructions as needed depending on the register file size. Next, local barriers are inserted such that every load following a store to the same address has at least one barrier in between. Finally, from these instructions, the decoupled streams are generated by assigning the instructions to their respective type of stream, while preserving the order.}

\textbf{Performance impact.} Due to the data prefetching enabled by the decoupled streams, the \name{} achieves 1.8$\times$ speedup over an in-order version of \name{} that uses the coupled instructions (over the benchmarks described in \S \ref{sec:experiments}).

\begin{figure}[!t]
\centering
\includegraphics[trim={0cm 0cm 0cm 0cm} , clip, width=\columnwidth]{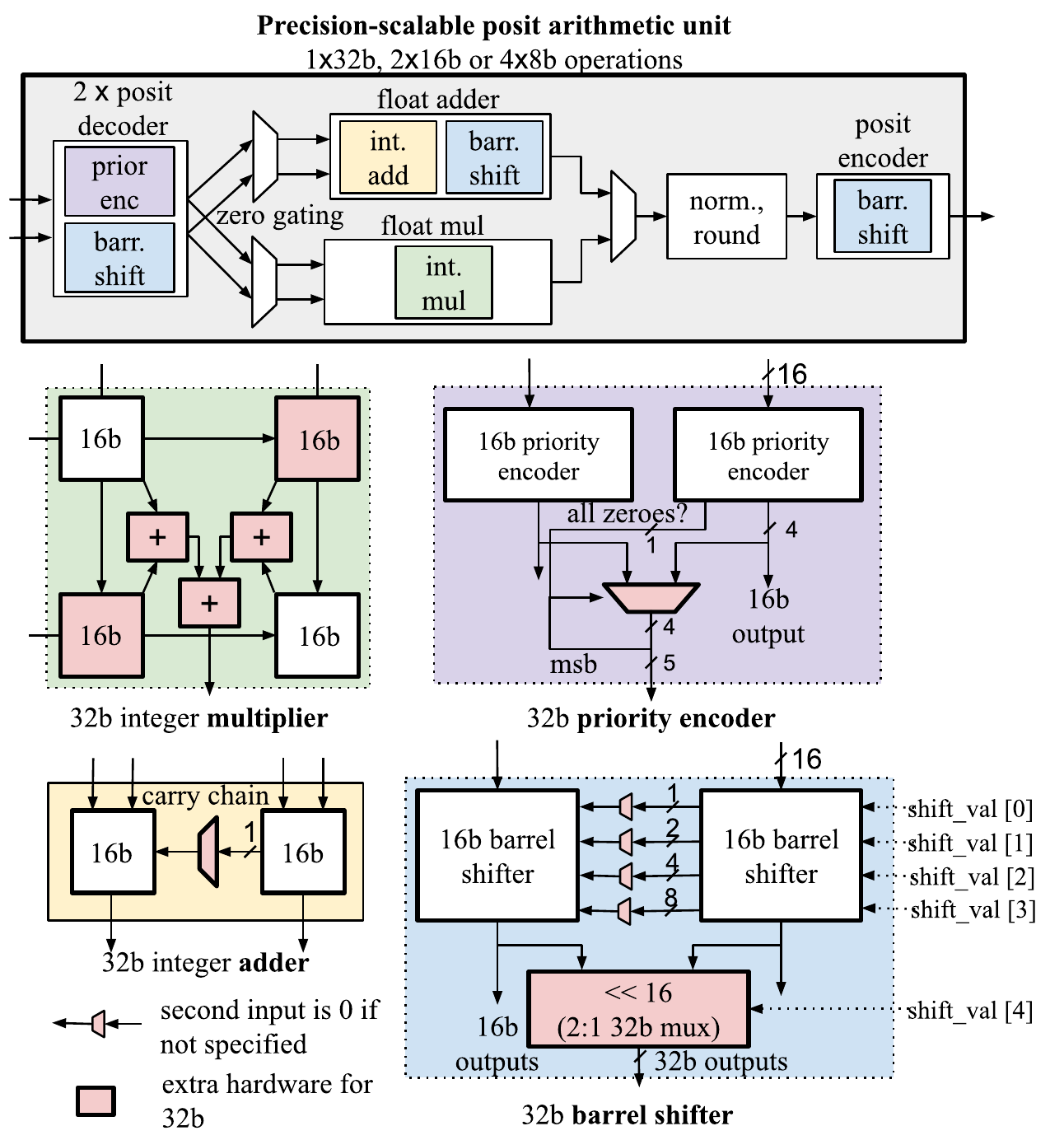}
\caption{Posit arithmetic unit with precision-scalable subunits}%
\label{fig:posit_architecture}
\end{figure}%

\section{Precision-scalable posit\textsuperscript{TM} unit} \label{sec:posit_unit}


\textbf{Application dependent precision-scalability}. DAGs from different applications can have widely varying precision requirements. For example, the probabilistic circuit (also called sum-product network) \cite{choi2020probabilistic,poon2011sum}, one of our irregular workloads, is used in machine learning for inference and reasoning under uncertainty. 
It can be used for safety-critical applications like autonomous navigation \cite{8967568} demanding highly accurate computation, but can also be deployed for simpler applications like human activity classification (running, sitting, etc.) \cite{galindez2019towards,shah2019problp}, which can tolerate some mispredictions.
To meet such diverse requirements, PEs are equipped with precision-scalable arithmetic units that can perform 1$\times$32b, 2$\times$16b or 4$\times$8b operations in a single cycle, enabling batch execution based on the application requirements.

\textbf{Posit\textsuperscript{TM} representation}. The \name{} uses the posit\textsuperscript{TM} \cite{gustafson2017beating} instead of the floating-point representation to enable lower-precision operations wherever possible. The posit representation (fig. \ref{fig:post_representation}) allows trading accuracy for range at runtime. To do this, it uses a new \textit{regime} field, whose length dynamically varies depending on the magnitude of the number, allowing to simply reduce the length of the other fields instead of under/overflowing. The posit representation is specified as a tuple ${<}L$, \textit{es}$>$ where $L$ is the total length and \textit{es} is the maximum length of the exponent field.

\textbf{Custom posit}. The standard posit representation proposed by Gustafson \cite{gustafson2017beating} aims to have higher precision than floats around the value of 1.0, while sacrificing precision for small and large numbers (fig.~\ref{fig:posit_accuracy_impact}(b)). While this design choice might be suitable for some applications, our target applications are characterized by a very large dynamic range and sensitivity to approximations across this range, requiring us to take a different approach.
In \name{}, higher \textit{es} is used (fig. \ref{fig:posit_accuracy_impact}(a)) such that the custom posit has a similar precision to float around 1.0, but a more gradual precision degradation for small/large values, greatly increasing the representable range of values. Fig. \ref{fig:posit_accuracy_impact}(b) demonstrates this gradual, tapered degradation of precision.

\textbf{Application accuracy with custom posit.} \rev{To quantify the impact of lower-precision computation on an application's accuracy, the arithmetic operations ($+$ and $\times$) in posit and float representations are prototyped with software models in Python.} Fig. \ref{fig:posit_accuracy_impact}(c) and (d) shows the impact of lower precision float, standard posit, and custom posit on different applications. Fig. \ref{fig:posit_accuracy_impact}(c) shows classification accuracy on machine learning tasks from UCI repository \cite{Dua:2019} with probabilistic circuit DAGs, while fig. \ref{fig:posit_accuracy_impact}(d) shows the relative error during iterative solutions of sparse matrix triangular systems on SuiteSparse matrices \cite{DBLP:journals/toms/DavisH11}. These 
benchmarks are described in detail in 
\S \ref{sec:experiments}. The results demonstrate that the custom posit outperforms the standard posit and float counterparts of the same lengths.

\textbf{Hardware design.} A posit operator at its core needs a float operator, since, for a given value of regime, posit behaves like a float. Hence, a posit operation is strictly costlier than a float operation (ignoring the exceptions of IEEE float). The arithmetic unit contains posit-format decoders and encoders, shared among float adder and multiplier (fig.~\ref{fig:posit_architecture}). The decoder finds the length of the regime field (with a priority encoder) and aligns the rest of the fields accordingly (with a barrel shifter) for addition and multiplication. The encoder aligns the output (with a barrel shifter) according to the output regime.

\begin{table}[!t]
\centering
\caption{Posit unit area and power breakdown}
\begin{tabular}{lccccc}
\toprule
\multirow{2}{*}{} & \multicolumn{2}{c}{\textbf{Area}} & & \multicolumn{2}{c}{\textbf{Power}}  \\
 & $\times 10^3$ $\mu$m2 & \% & & $\mu$W & \% \\ \midrule
Decoders & 1.2 & 17 & & 166 & 19  \\
Float add and mul & 4.3 & 57 & & 484 & 56  \\
Normalize and round & 1.4 & 18 & & 142 & 17  \\
Encoder & 0.6 & 8 & & 64 & 8  \\ \midrule
\textbf{Precision-scalable posit unit} & 7.5 &  & & 856 &  \\ 
\bottomrule
\end{tabular}%
\label{tab:posit_area_power_breakdown}
\end{table}

\begin{table}[!t]
\centering
\caption{\rev{Performance comparison with an state-of-the-art posit unit}}
\begin{tabular}{lcccc}
\toprule
 & \multicolumn{3}{c}{This work} & PACoGen \cite{jaiswal2019pacogen} \\ \midrule
Operating mode & 32b & 16b & 8b &  32b \\ 
Area ($\times 10^3$ $\mu$m2) & \multicolumn{3}{c}{7.5} & 5.7 \\
Energy (pJ/op) & 3.05 & 1.33 & 0.57 & 2.08\\
Throughput (ops/cycle) & 1 & 2 & 4 & 1 \\
\bottomrule
\end{tabular}%
\label{tab:posit_PACOGEN_comparison}
\end{table}

All the blocks in the arithmetic units support precision scalability to perform 1$\times$32b, 2$\times$16b or 4$\times$8b operations. This runtime scalability is novel, and not available in other posit hardware generators \cite{chaurasiya2018parameterized, jaiswal2019pacogen, tiwari2021peri}. Fig. \ref{fig:posit_architecture} shows how 32b building blocks are constructed from two 16b blocks, which in turn are made of 8b blocks. The posit unit consumes 1.8$\times$ the area and power of a float counterpart (Table \ref{tab:posit_area_power_breakdown}), but enables 8b or 16b operations as discussed earlier. 
\rev{Table \mbox{\ref{tab:posit_PACOGEN_comparison}} reports comparison with PACoGen \mbox{\cite{jaiswal2019pacogen}}, a state-of-the-art posit unit generator, which consumes 
0.76$\times$ and 0.68$\times$ area and energy at 32b, respectively, due to the overhead of precision-scalability in our unit. On the other hand, for applications requiring only 8b precision, the 32b PACoGen unit consumes 3.7$\times$ the energy per operation of the precision-scalable unit.}

\section{Experiments} \label{sec:experiments}
\begin{figure}[!t]
\centering
\includegraphics[trim={0cm 0cm 0cm 0cm} , clip, width=0.9\columnwidth]{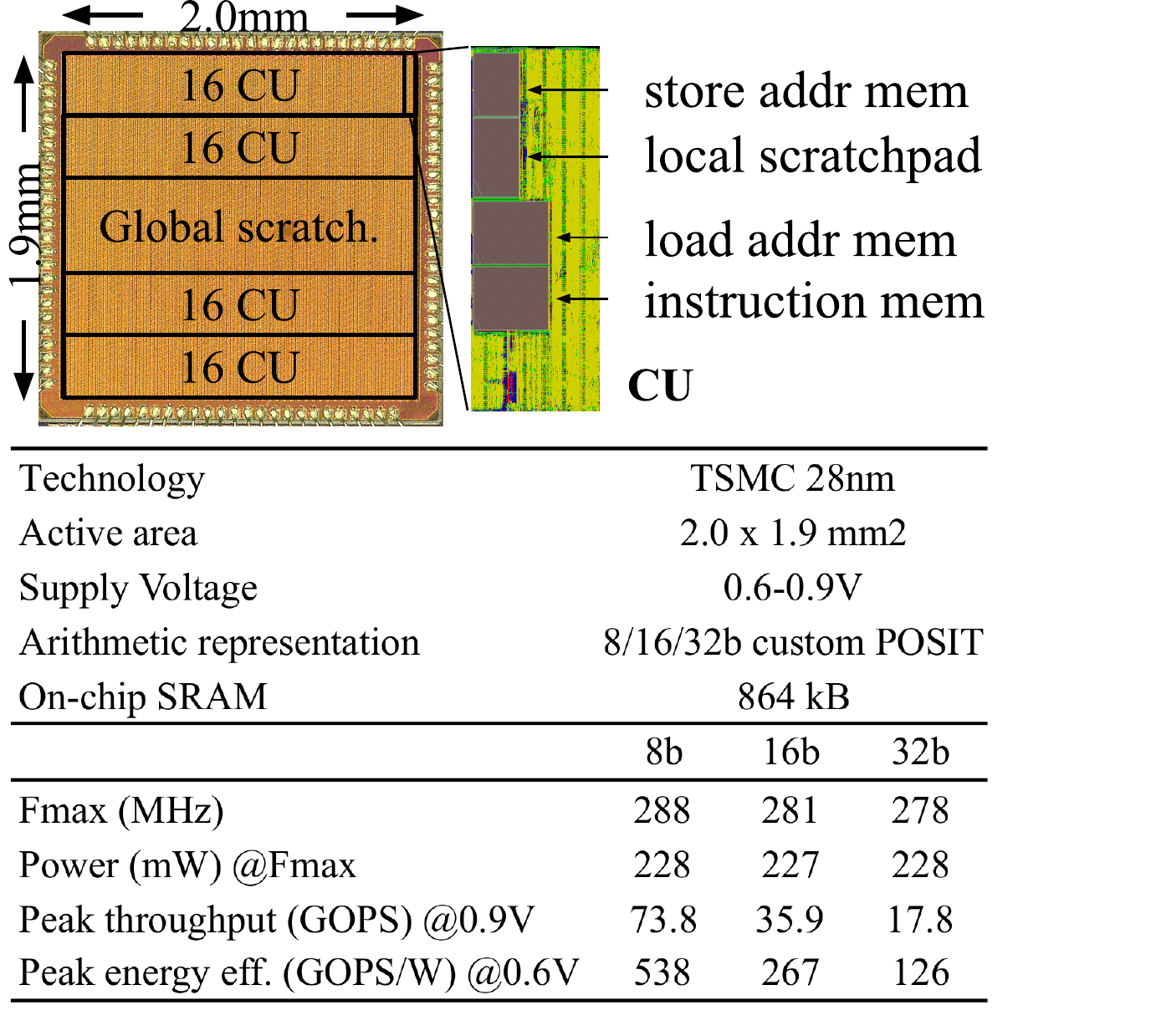}
\caption{Chip micrograph and specifications}%
\label{fig:chip_micrograph}
\end{figure}

\name{} is taped-out in TSMC 28nm technology with an active area of 2.0$\times$1.9mm$^2$ (fig. \ref{fig:chip_micrograph}). 
Table \ref{tab:full_area_power_breakdown} shows the post-layout area and power breakdown of the chip (estimated after activity annotation). The memories occupy half of the area and consume 37\% of the power. The asymmetric crossbar consumes 13\% power, which would have been two times costlier if a conventional symmetric crossbar had been used. 

\begin{table}[!t]
\centering
\caption{Post-layout area and power breakdown}
\begin{tabular}{lccccc}
\toprule
\multirow{2}{*}{} & \multicolumn{2}{c}{\textbf{Area}} & & \multicolumn{2}{c}{\textbf{Power}}  \\
 & mm$^2$ & \% & & mW & \% \\ \midrule
64 Compute Units: &  &  &  &  &  \\
\hspace{5mm}PEs: &  &  &  &  &  \\
\hspace{10mm}Posit units & 0.48 & 13 & & 54.8 & 24  \\
\hspace{10mm}Rest of PE & 0.50 & 13 & & 37.4 & 16 \\
\hspace{5mm}Local scratchpads & 0.27 & 7 & & 13.6 & 6 \\
\hspace{5mm}Instr. mem, ld/st addr. mem & 1.18 & 31 & & 50.2 & 22 \\
\hspace{5mm}Rest of CU & 0.24 & 6 & &22.6 & 10 \\
Global scratchpads & 0.54 & 14 & & 21.0 & 9 \\
Crossbar & 0.19 & 5 & & 29.3 & 13 \\
Rest & 0.40 & 11 &  &  &  \\ \midrule
\textbf{\name{}} & 3.8 &  & & 228.9 &  \\ \bottomrule
\end{tabular}%
\label{tab:full_area_power_breakdown}
\end{table}

\begin{figure}[!t]
\centering
\includegraphics[trim={0cm 0.4cm 0cm 0cm} , clip, width=\columnwidth]{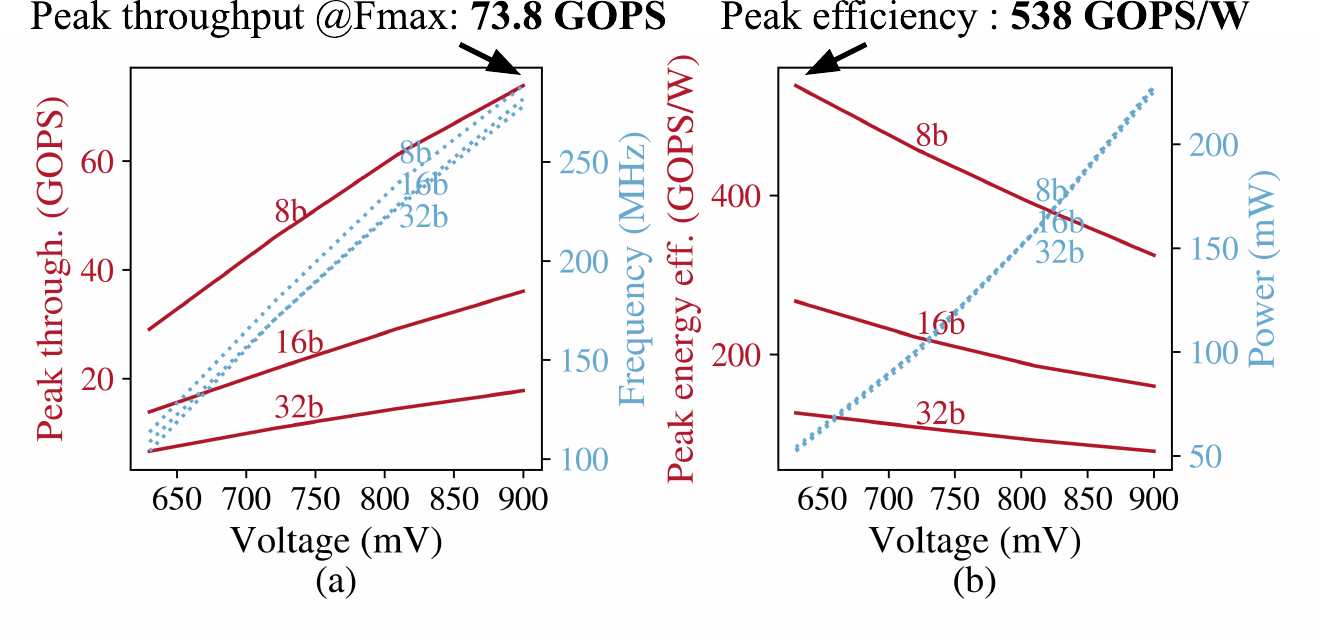}
\caption{Peak-performance scaling with voltage and precision}%
\label{fig:voltage_scaling}
\end{figure}%

\subsection{Peak performance and voltage scaling} \label{sec:peak_performance}
The chip's electrical performance is measured by scaling the voltage from the nominal 0.9V to 0.6V. The chip can operate at the maximum frequency of 288MHz at 0.9V and 8b precision, with a peak throughput of 73.8 GOPS (fig. \ref{fig:voltage_scaling}(a)), and at the peak energy-efficiency of 538 GOPS/W at 0.6V (fig. \ref{fig:voltage_scaling}(b)). \rev{The top 50 critical paths are in the posit arithmetic unit and the crossbar. Both the blocks can be pipelined to increase clock frequency further, but would have a negative impact on instruction scheduling due to higher posit unit latency, and would induce a potential throughput tradeoff due to increased access latency of the global scratchpad.}


\renewcommand{\arraystretch}{1}
\begin{table}[]
\caption{Statistics of the benchmarked DAGs}
\label{tab:workload_details}
\centering
\begin{tabular}{ccccc}
\toprule
Application & Workload & \begin{tabular}[c]{@{}c@{}}Nodes\\ (n)\end{tabular} & \begin{tabular}[c]{@{}c@{}}Longest path \\ length (l)\end{tabular} & \begin{tabular}[c]{@{}c@{}}Parallelism\\ (n/l)\end{tabular} \\ \midrule
\multirow{9}{*}{\begin{tabular}[c]{@{}c@{}}Probabilistic\\ Circuits (PC)\end{tabular}}
 & mnist      & 10414  & 26 & 400 \\
 & nltcs      & 13627  & 27 & 504 \\
 & msnbc      & 47334  & 28 & 1690 \\
 & bnetflix   & 55007  & 53 & 1037 \\
 & ad         & 66819  & 93 & 718 \\
 & bbc        & 77457  & 92 & 841 \\
 & c20ng      & 80962  & 81 & 999 \\
 & kdd        & 98211  & 54 & 1818 \\
 & baudio     & 121263 & 70 & 1732 \\
 & pumsbstar & 149662  & 82 & 1825 \\

 \midrule
\multirow{11}{*}{\begin{tabular}[c]{@{}c@{}}Sparse Matrix \\ Triangular Solves \\ (SpTRSV)\end{tabular}} 
& tols4000 & 5978   & 52   & 114\\
& bp\_200   & 8406   & 139  & 60 \\
& west2021 & 10159  & 136  & 74 \\
& qh1484   & 11298  & 237  & 47 \\
& sieber   & 22768  & 242  & 94 \\
& gemat12  & 74199  & 778  & 95 \\
& dw2048   & 79240  & 929  & 85 \\
& orani678 & 114275 & 634  & 180\\
& pde2961  & 140303 & 1357 & 103\\
& blckhole & 150876 & 1264 & 119\\
\bottomrule
\end{tabular}%
\end{table}

\subsection{Workloads}\label{sec:workloads}
The performance of the chip is benchmarked with compute DAGs from two different types of workloads listed in table \ref{tab:workload_details}. \rev{The \name{} compiler takes as input a DAG in any of the popular graph formats (i.e. all formats supported by the NetworkX package \mbox{\cite{hagberg2008exploring}})
and generates an execution binary that can be directly programmed to \name{}.}
The experiments are performed with the DAGs that fit in the on-chip data memory (global and local scratchpads).
The programming of memories, with an FPGA via a slow chip I/O interface, is not included in the throughput results. 

\textbf{Probabilistic circuits (PC).} Probabilistic circuits (also called sum-product networks) are used in machine learning for reasoning, robust inference under uncertainty, and safety-critical tasks 
    \cite{DBLP:conf/icml/StelznerPK19,8967568,galindez2019towards, thoma2021recowns}. Note that the term \textit{circuit} is not used in the sense of VLSI circuits. A PC is an irregular DAG in which the nodes are either sum or product operations.
    The experiments are performed on a standard benchmark of density-estimation applications
    from \cite{DBLP:conf/uai/LiangBB17}.
    
\textbf{Sparse matrix triangular solves (SpTRSV).} Solving a matrix triangular system is a fundamental operation in linear algebra, used in various applications like robotics, optimization, autonomous navigation, etc. Real-world matrices usually turn out to be highly sparse and the resulting compute DAG highly irregular. Benchmarking is done on matrices of varying sizes from the SuiteSparse benchmark of real applications \cite{DBLP:journals/toms/DavisH11}.
\subsection{Throughput scaling with different active CUs}
\begin{figure}[!t]
\centering
\includegraphics[trim={0cm 0.5cm 0cm 0cm} , clip, width=0.7\columnwidth]{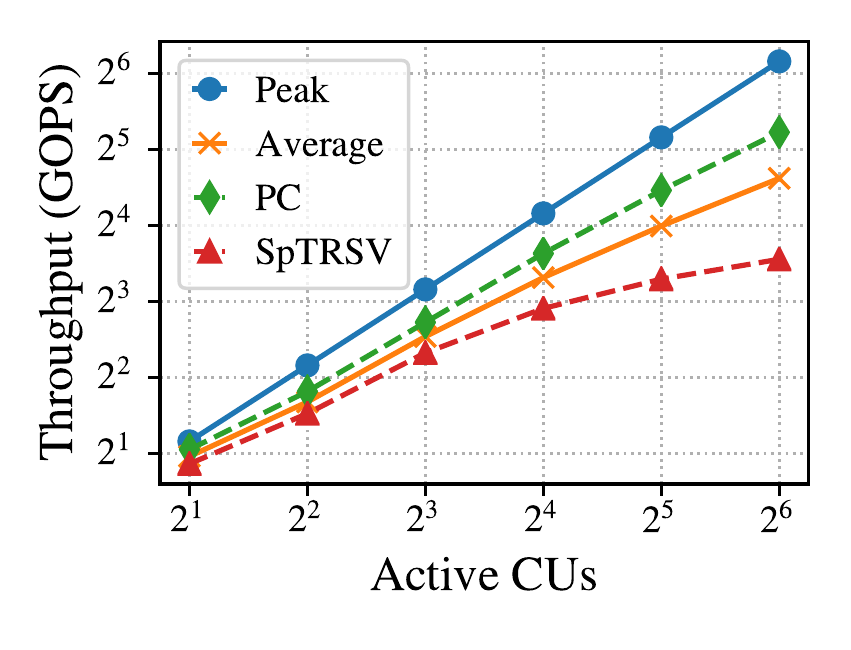}
\caption{Scaling of throughput with increasing active CUs at 8b precision}%
\label{fig:cu_scaling}
\end{figure}

The throughput of \name{} is measured by keeping a different number of CUs active to evaluate the parallelization effectiveness for increasing parallel CUs (fig. \ref{fig:cu_scaling}). The PC throughput scales better than the SpTRSV because of higher DAG parallelism (table \ref{tab:workload_details}). Apart from DAG parallelism, other factors that lower average throughput compared to the peak are: (1) the impact of barriers on data-prefetching, and (2) global scratchpad access conflicts. 
Overall, the average throughput is 2.8$\times$ lower than the peak for 64 CUs.

\subsection{Comparison with CPU and GPU}
Since there is no previous silicon-proven chip targeting highly irregular DAGs, this experiment is designed to compare \name{}'s performance with state-of-the-art CPU and GPU implementations. The details of the platforms are as follows. \textbf{\name{}:} The results are for 64 active CUs operating at 278MHz and 32b precision. \textbf{CPU:} An Intel(R) Xeon Gold 6154 CPU operating at 3GHz is used for comparison. For PC, a standard Julia-based library called Juice \cite{dang2021juice} (\textbf{CPU-JUICE}) and an highly-optimized OpenMP-based implementation \cite{shah2021graphopt} (\textbf{CPU-OMP}) are used for comparison. The SpTRSV performance is evaluated with the standard Intel Math Kernel Library (MKL v2021.1). The programs are compiled with GCC v4.8.5 compiler, \texttt{-Ofast} flag, and OpenMP v3.1.

\begin{figure}[!t]
\centering
\includegraphics[trim={0cm 0cm 0cm 0cm}, clip, width=\columnwidth]{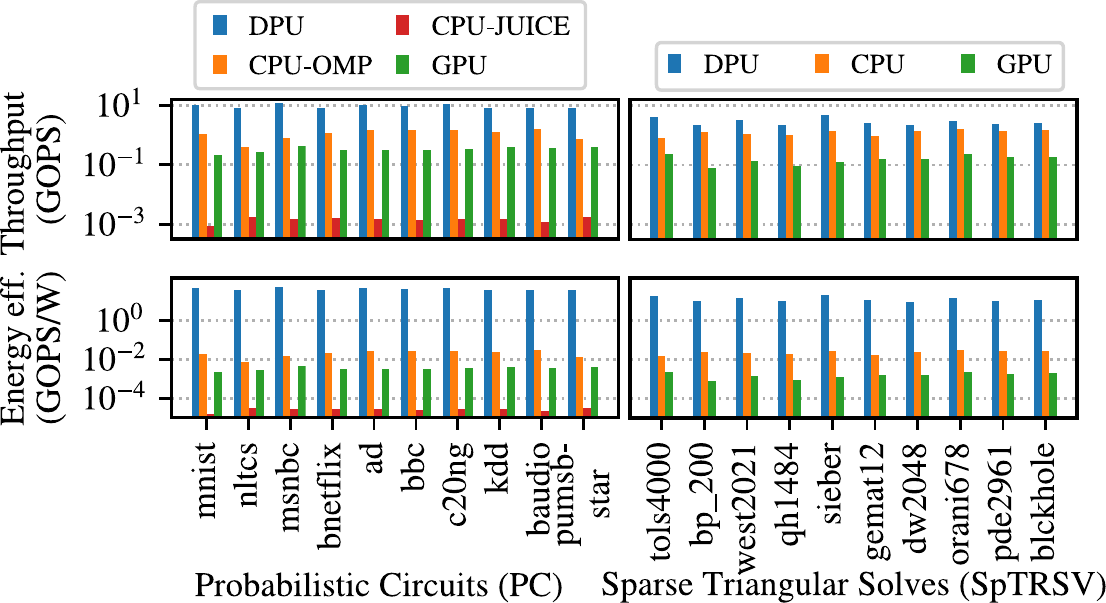}
\caption{Performance comparison. The \name{} operating point is 0.9V and 32b.}%
\label{fig:throughput_for_workloads}
\end{figure}%

\begin{table}[!t]
\centering
\caption{Performance comparison with other platforms.}
\label{tab:power_energy_eff}
\begin{tabular}{lccc}
\toprule
 & \name{}$^*$ & CPU & GPU \\ \midrule
Technology  & 28nm & 22nm & 12nm \\
Frequency (GHz) & 0.28 & 3 & 1.35 \\
Arithmetic representation & custom posit & float & float \\
Peak throughput (GOPS) & 17.8 & 3.4$\times$10$^3$ & 13.5$\times$10$^3$ \\
Avg. throughput (GOPS) & 6.2 & 1.2 & 0.3 \\
Power (W) & 0.23 & 55 & 98 \\
Avg. energy efficiency (GOPS/W) & 27 & 0.02 & 0.003 \\ 
Workloads & \multicolumn{3}{c}{PC and SpTRSV DAGs} \\
\bottomrule
\multicolumn{4}{l}{$^*$\name{} operating point is 0.9V and 32b precision}
\end{tabular}%
\end{table}

\textbf{GPU:} The GPU baseline is evaluated with an RTX 2080Ti GPU operating at 1.3GHz, and the code compiled with the CUDA v10.2.89 compiler. For PC, an efficient CUDA code described in \cite{shah2020acceleration} is used for benchmarking. For SpTRSV,  the \texttt{cusparseScsrsv\_solve()} function from the standard cuSPARSE library \cite{naumov2010cusparse, naumov2011parallel} is used. For a fair comparison, the memory copy time from the host to GPU is not considered.

Fig. \ref{fig:throughput_for_workloads} and table \ref{tab:power_energy_eff} summarizes the comparison results. All the platforms show higher performance for PC than SpTRSV due to the higher parallelism (table \ref{tab:workload_details}). Juice performs considerably slower than the OpenMP counterpart, despite using the same number of CPU cores. Overall, the \name{} outperforms the CPU, which in turn beats the GPU. The \name{} achieves an average throughput of 6.2 GOPS, a speedup of  5.1$\times$ and 20.6$\times$ compared to the CPU and GPU, at an average efficiency of 27 GOPS/W at 32b precision, showing the effectiveness of the specialized \name{} architecture for irregular DAGs.

\subsection{\name{}'s performance for a regular DAG} \label{sec:experiment_regular_dag}
\rev{\name{}'s performance for \textit{regular} DAGs would be an interesting result, quantifying the effectiveness of asynchronous CUs for a regular workload. Hence, as an additional experiment, performance is benchmarked for a regular DAG of dense matrix-vector multiplication (GEMV). For a 128x128 matrix, \name{} achieves a throughput of 17.5 GOPS at a utilization of 97.5\%, while consuming 414mW at  0.9V, 0.28GHz and 32b precision,
resulting in an efficiency of 42 GOPS/W. This shows that \name{} can achieve near-peak throughput for a regular DAG, although with an inefficiency that separate instructions and load/store addresses are used for every CU due to the absence of SIMD support. For reference, the CPU achieves 7.4 GOPS and 0.14 GOPS/W with the Intel MKL GEMV function \texttt{cblas\_sgemv()}.
}

\section{Related work} \label{sec:related_works}

Neural-network processors exploiting sparsity like \cite{parashar2017scnn, 9310233, 8890710,he2020sparse, kwon2018maeri} have special hardware support to handle irregularity resulting from the sparsity. However, sparsity in our DAG workloads is typically more than 99\%, significantly higher than NN sparsity (less than 70-80\%). As a result, sparse NNs exhibit higher compute-to-memory fetch ratios and some repetitive structures that can be exploited, e.g., by using a systolic array of PEs, while the \name{} needs different architectural techniques due to the ultra-high sparsity.

In recent years, architectures like \cite{gui2019survey, yao2018efficient, addisie2018heterogeneous,ham2016graphicionado,li2018graphia} have been proposed for general graph-analytic workloads like PageRank, breadth-first search, single-source shortest paths, etc. The key difference is that these architectures work well when a significant portion of the graph nodes are \textit{active}, while in compute DAGs only the nodes in a DAG layer are active at a time due to the dependency-induced node ordering. In general graph analytics, the active nodes cannot be predicted at compile time, while they can be predicted for our target DAGs, which is heavily utilized in this work for reducing the number of barriers, aggressive data prefetching, etc. 

The sparse processing unit (SPU) \cite{dadu2019towards} uses hardware support for \textit{stream-joins} which is similar to our decoupled streams (\S \ref{sec:decoupled_streams}). However, \name{} uses a register-bank based PE for local data reuse, while SPU uses a coarse-grained reconfigurable dataflow array (CGRA). Such a dataflow array can be fully utilized when there are frequently recurring patterns in the DAG for which the array can be reconfigured, but irregular DAGs typically lack a single repetitive pattern. Further, the SPU consumes 16W (simulated) as opposed to \name{}'s 0.23W.

Overall, \name{} is novel in targeting DAG applications with high sparsity-induced irregularity, with specialized features like hardware-supported synchronization, decoupled streams-based execution, and application-dependent posit arithmetic.

\section{Conclusion} \label{sec:conclusion}
This paper proposed \name{}, a processor designed for energy-efficient parallel execution of irregular DAGs. 
The \name{} is equipped with 64 parallel compute units (CU), each executing a DAG subgraph independently. The CUs communicate via a high-bandwidth global scratchpad connected using a low-overhead asymmetric crossbar. Synchronization of CUs, frequently needed for DAGs, happens in a single cycle using a specialized hardware unit. The instructions of CUs are decoupled into multiple streams for overlapping execution, resulting in 1.8$\times$ speedup. For the arithmetic operations, the CUs are equipped with precision-scalable custom posit units that can perform low-precision batch inference depending on the application requirement. The \name{} is fabricated in 28nm technology, and benchmarked on irregular DAGs from probabilistic machine learning and sparse linear algebra. Measurement results show a mean speedup of 5.1$\times$ and 20.6$\times$ over state-of-the-art CPU and GPU implementations, with a peak performance of 73.8 GOPS and 538 GOPS/W. Thus, \name{} takes a step towards supporting emerging irregular DAG workloads in energy-constrained platforms.

\section*{Acknowledgment}

This work has been supported by the EU ERC project Re-SENSE under grant agreement ERC-2016-STG-715037,
and we acknowledge EUROPRACTICE MPW and design tool support, and support from Intel.


%



\clearpage

\ifCLASSOPTIONcaptionsoff
  \newpage
\fi



\bibliographystyle{IEEEtran}
\bibliography{IEEEabrv,ref.bib}

\clearpage


%



\newpage
%

\begin{IEEEbiography}[{\includegraphics[width=1in,height=1.25in,clip,trim={4cm 17cm 4cm 0cm},keepaspectratio]{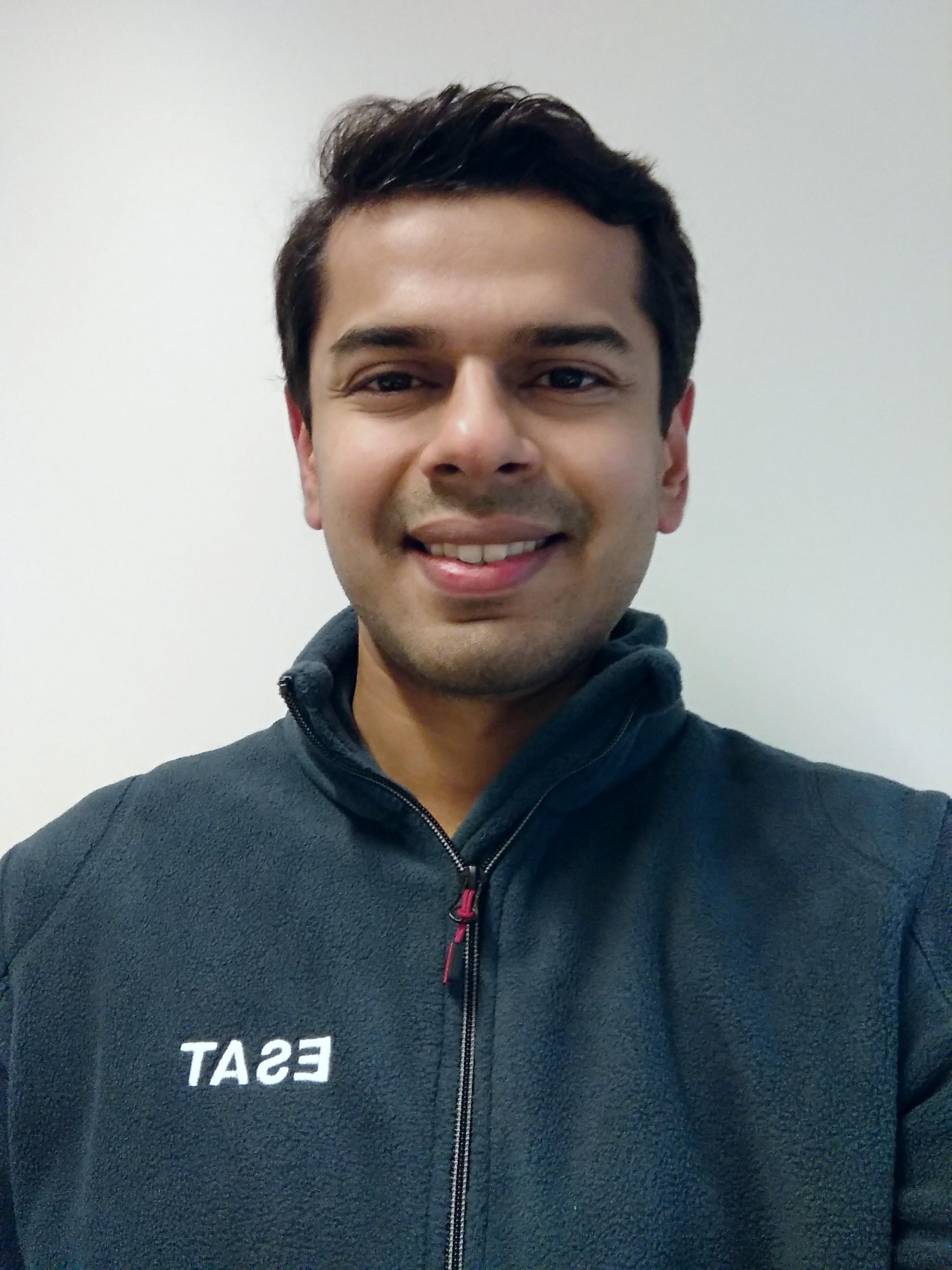}}]{Nimish Shah}
Nimish Shah is pursuing his Ph.D. degree at the MICAS laboratories of the EE department of KU Leuven, Belgium. His research focuses on hardware-software co-design, embedded machine learning, irregular graph processing, approximate computing, and low-power digital VLSI. He received an M.Tech. degree in Electronic Systems Engineering from the Indian Institute of Science, Bangalore, in 2016. In 2016-17, he worked with Nvidia, Bangalore, where he was involved in energy-efficient memory (de)compression hardware design for GPU. He has served as a reviewer for TVLSI and JETCAS. Nimish is a recipient of the departmental Gold Medal for excellence in master’s studies at IISc.
\end{IEEEbiography}

\begin{IEEEbiography}[{\includegraphics[width=1in,height=1.25in,clip,keepaspectratio]{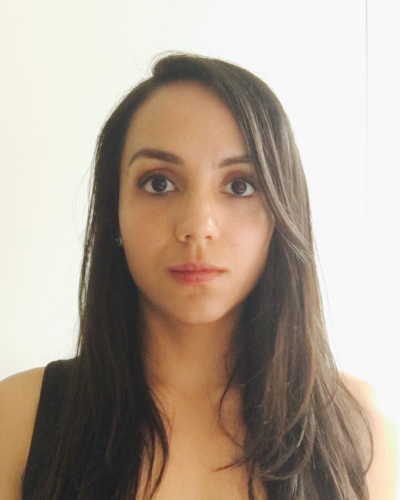}}]{Laura Isabel Galindez Olascoaga}
Laura Isabel Galindez Olascoaga received the M.Sc. degree in Systems and Control from TU Eindhoven, The Netherlands, in 2015, and the Ph.D. degree in Electrical Engineering from KU Leuven, Belgium, in 2020. Since February 2021, she has been a postdoctoral research scholar at the Berkeley Wireless Research Center in the Electrical Engineering and Computer Sciences Department of the University of California, Berkeley, USA. In 2014, she was a research intern at ASML, Veldhoven, The Netherlands; and in 2018, she was a visiting researcher at the Statistical and Relational Artificial Intelligence Lab of the University of California, Los Angeles. Her research interests include hardware-aware machine learning, tractable probabilistic models and Probabilistic Circuits, brain-inspired high-dimensional computing, neurosymbolic Artificial Intelligence and human-in-the-loop robot control and learning.
\end{IEEEbiography}

\begin{IEEEbiography}[{\includegraphics[width=1in,height=1.25in,clip,keepaspectratio]{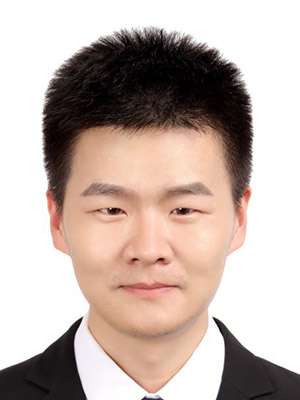}}]{Shirui Zhao}

Shirui Zhao received his B.S. degree from Northwestern Polytechnical University (NWPU) in 2012 and M.S. degree from the University of Chinese Academy of Sciences (UCAS) in 2015, respectively. He is currently pursuing a Ph.D. degree at ESAT-MICAS, KU Leuven, with a focus on the area of low-power probabilistic reasoning hardware design.
His research interests span machine learning, chip architecture, and low-power circuit design.
\end{IEEEbiography}

\begin{IEEEbiography}[{\includegraphics[width=1in,height=1.25in,clip, keepaspectratio]{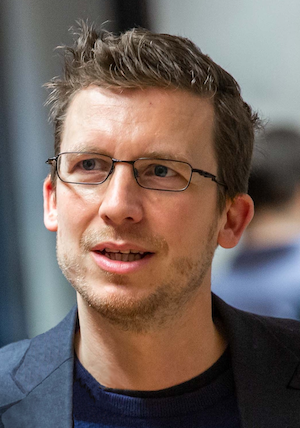}}]{Wannes Meert}
Wannes Meert received his degrees of Master of Electrotechnical Engineering, Micro-electronics (2005), Master of Artificial Intelligence (2006) and Ph.D. in Computer Science (2011) from KU Leuven. He is an IOF research manager in the DTAI section at KU Leuven. His work is focused on applying machine learning, artificial intelligence and anomaly detection technology to industrial application domains with various industrial and academic partners.
\end{IEEEbiography}

\begin{IEEEbiography}[{\includegraphics[width=1in,height=1.25in,clip,keepaspectratio]{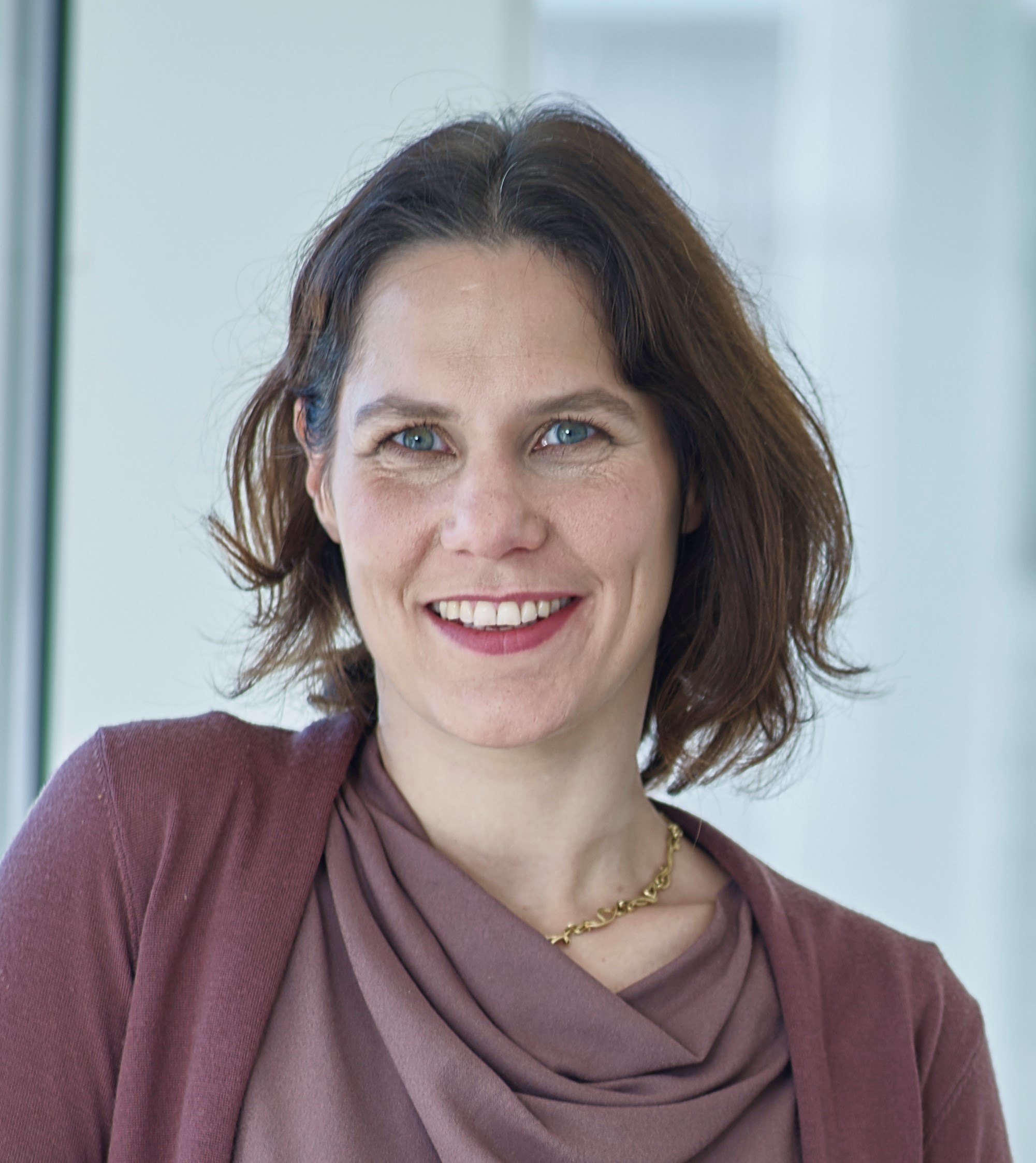}}]{Marian Verhelst}
Marian Verhelst is a full professor at the MICAS laboratories of the EE Department of KU Leuven. Her research focuses on embedded machine learning, hardware accelerators, HW-algorithm co-design and low-power edge processing. Before that, she received a PhD from KU Leuven in 2008, was a visiting scholar at the BWRC of UC Berkeley in the summer of 2005, and worked as a research scientist at Intel Labs, Hillsboro OR from 2008 till 2011. Marian is a topic chair of the DATE and ISSCC executive committees, TPC member of VLSI and ESSCIRC  and was the chair of tinyML2021 and TPC co-chair of AICAS2020. Marian is an IEEE SSCS Distinguished Lecturer, was a member of the Young Academy of Belgium, an associate editor for TVLSI, TCAS-II and JSSC and a member of the STEM advisory committee to the Flemish Government. Marian currently holds a prestigious ERC Starting Grant from the European Union, was the laureate of the Royal Academy of Belgium in 2016, and received the André Mischke YAE Prize for Science and Policy in 2021.
\end{IEEEbiography}







\end{document}